\documentclass[natbib]{svjour3}
\smartqed  % flush right qed marks, e.g. at end of proof
\usepackage{graphicx}
\usepackage{mathptmx}      % use Times fonts if available on your TeX system
\usepackage{aps-bibstyle}  % use this style if you don't use BibTeX.

\journalname{Space Science Reviews}

\begin{document}

\title{Self Consistent Models of the Solar Wind}
\titlerunning{Self Consistent Solar Wind Models}

\author{Steven R. Cranmer}
\authorrunning{S. R. Cranmer}

\institute{S. R. Cranmer \at
Harvard-Smithsonian Center for Astrophysics,
60 Garden Street, Cambridge, MA 02138, USA \\
Tel.: +1-617-495-7271 \\
\email{scranmer@cfa.harvard.edu}}

\date{Received: March 26, 2010 / Accepted: July 5, 2010}

\maketitle
\begin{abstract}
The origins of the hot solar corona and the supersonically expanding
solar wind are still the subject of much debate.
This paper summarizes some of the essential ingredients of
realistic and self-consistent models of solar wind acceleration.
It also outlines the major issues in the recent debate over what
physical processes dominate the mass, momentum, and energy balance
in the accelerating wind.
A key obstacle in the way of producing realistic simulations of the
Sun-heliosphere system is the lack of a physically motivated way
of specifying the coronal heating rate.
Recent models that assume the energy comes from Alfv\'{e}n waves that
are partially reflected, and then dissipated by magnetohydrodynamic
turbulence, have been found to reproduce many of the observed
features of the solar wind.
This paper discusses results from these models, including detailed
comparisons with measured plasma properties as a function of
solar wind speed.
Some suggestions are also given for future work that could answer
the many remaining questions about coronal heating and solar
wind acceleration.
\end{abstract}

\section{Introduction}
\label{sec:1}

It has been known since the early part of the twentieth century
that the temperature in the Sun's outer atmosphere undergoes a
rapid inversion, from a relatively cool ($T < 10^{4}$ K) photosphere
and chromosphere to a hot and ionized ($T > 10^{6}$ K) corona.
Since the dawn of the space age, coronal heating has also been known
to be linked to the acceleration of a plasma outflow that
reaches speeds of 300--800 km s$^{-1}$ in interplanetary space.
In the years since the twin problems of coronal heating and solar
wind acceleration were formulated, many physical processes have been
suggested to be responsible.
Only a small fraction of the mechanical energy in the Sun's
sub-photospheric convection zone needs to be converted to heat in order
to power the corona.
However, it has proved exceedingly difficult to distinguish between
competing theoretical models using existing observations.
Recent summaries of these problems and controversies have been
presented by, e.g., \citet{As06}, \citet{Kl06}, \citet{Zu07},
\citet{Ho08}, and \citet{Cr09}.

Most of the proposed physical models can be grouped into two broad
paradigms.
As will be seen below, the basic difference between these paradigms
concerns the overall topological connectivity of the magnetic flux
tubes that feed the solar wind:

\begin{enumerate}
\item
If solar wind flux tubes are open to interplanetary space---and if
they remain open on timescales comparable to the time it takes plasma
to accelerate into the corona---then the main sources of energy must
be injected at the footpoints of the flux tubes.
Thus, in {\em wave/turbulence-driven (WTD) models,}
the convection-driven jostling of the flux-tube footpoints
is assumed to generate wave-like fluctuations that propagate up into
the extended corona.
These waves (usually Alfv\'{e}n waves) are often proposed to
partially reflect back down toward the Sun, develop into strong
magnetohydrodynamic (MHD) turbulence, and dissipate gradually.
These models have been shown to naturally produce realistic fast
and slow wind conditions with wave amplitudes of the same order of
magnitude as those observed in the corona and heliosphere
\citep{Ho86,Ve91,WS91,Mt99,SI06,CvB07,Wa09,VV10,MS10}.
\item
Near the Sun, all open magnetic flux tubes are observed to exist in
the vicinity of closed loops that are evolving on a wide range of
spatial and time scales in a complex ``magnetic carpet'' \citep{TS98}.
Thus, it is natural to propose a class of
{\em reconnection/loop-opening (RLO) models,} in which
solar wind flux tubes are assumed to be influenced by magnetic
reconnection with closed-field regions that are continuously emerging,
fragmenting, and being otherwise jostled by convection.
In these models the mass, momentum, and energy of the solar wind is
input from loops of varying properties in the low corona.
Some have suggested that RLO-type energy exchange primarily occurs
on small, supergranular scales \citep{Ax92,Fi99,Fi03,SM03}.
However, other models have been proposed in which the reconnection
occurs in and between large-scale coronal streamers
\citep{Ei99,SN04,An10}.
\end{enumerate}
In the interest of brevity, this paper does not review the wider
range of empirically based solar wind models that do not contain
self-consistent coronal heating physics.
These models are often quite sophisticated in their treatment of
multi-fluid \citep[e.g.,][]{Hn97,Li03,LE05} or multi-dimensional
\citep[e.g.,][]{Ev03,Ev04,Ro03,Ri06,Nk09} effects.
However, their use of {\em ad hoc} rates of heating and
momentum deposition places them in a separate category from the
self-consistent models discussed here.

\section{Essential Ingredients}
\label{sec:2}

Before reviewing the results of WTD or RLO models, it is useful to
summarize what kinds of ``physics inputs'' are necessary to build
a self-consistent model of solar wind acceleration.

The first models of the solar wind were not self-consistent, but
they contained many valuable insights that led the way to future
improvements.
\citet{P58} found that a constant coronal temperature of order
$10^6$ K (i.e., roughly consistent with strong radial electron
conduction) provides enough of an outward gas pressure gradient
force to overcome gravity and produce a natural transition from a
subsonic (bound, negative energy) state near the Sun to a
supersonic (escaping, positive energy) state in interplanetary space.
Throughout the 1960s, new models with different prescribed coronal
temperatures $T(r)$ explored the parameter space of possible
time-steady solar wind solutions.
Some of these models included temperatures consistent with
polytropic equations of state (i.e., $P \propto \rho^{\gamma}$),
and \citet{HA70} found that $\gamma \leq 1.5$ is necessary for an
accelerating solar wind.
Also, \citet{SH66} solved two-fluid ($T_{p} \neq T_{e}$) energy
equations with heat conduction, and found that some kind of ``extra''
energy addition is needed in the corona to heat the plasma to the
temperatures seen at 1 AU.

The order-of-magnitude amount of coronal heating that is needed to
produce the solar wind can be estimated from an approximate version of
the internal energy conservation equation \citep[see, e.g.,][]{Le82}.
The time-steady version of this equation,
\begin{equation}
  \nabla \cdot \left[ {\bf F}_{\rm heat} + {\bf F}_{\rm cond} +
  \rho {\bf u} \left( \frac{u^2}{2} + \frac{5P}{2\rho} -
  \frac{GM_{\odot}}{r} \right) \right] \, = \, Q_{\rm rad}
  \,\, ,
\end{equation}
describes the balance between an imposed heat flux
(${\bf F}_{\rm heat}$), conduction along the magnetic field
(${\bf F}_{\rm cond}$), radiative losses ($Q_{\rm rad}$), and
fluxes of kinetic energy, enthalpy, and gravitational potential
energy (the three terms in parentheses).
When studying how the energy budget varies from the corona to 1 AU,
the dominant terms are the imposed heating, the kinetic energy flux,
and gravity.
Thus, \citet{HL95} found that one can estimate
\begin{equation}
  F_{\rm heat} \, = \, |{\bf F}_{\rm heat}| \, \approx \,
  \left( \rho u \right)_{\rm corona}
  \left( \frac{V_{\rm esc}^2}{2} + \frac{u_{\infty}^2}{2} \right)
\end{equation}
\citep[see also][]{LM99,SM03}.
Thus, if we can specify the mass flux in the corona, $\rho u$,
the escape velocity from the solar surface,
$V_{\rm esc} = (2GM_{\odot} / R_{\odot})^{1/2}$,
and the asymptotic, or terminal outflow speed, $u_{\infty}$,
we can estimate how much heat must be deposited in the corona.
Using typical values \citep{CvB07} for the fast solar wind
associated with coronal holes, $F_{\rm heat} \approx 8 \times 10^{5}$
erg cm$^{-2}$ s$^{-1}$.
For the slow solar wind associated with streamers and active regions,
$F_{\rm heat} \approx 3 \times 10^{6}$ erg cm$^{-2}$ s$^{-1}$.

In order to go beyond order-of-magnitude estimates, models
must begin to include self-consistent descriptions of specific
physical processes involved in the mass, momentum, and energy
conservation of accelerating plasma parcels.
There are at least five essential ingredients to such a self-consistent
picture of the solar wind:

\begin{enumerate}
\item
{\em Physically motivated coronal heating.}
The actual origin of the imposed heat flux $F_{\rm heat}$ (or
the equivalent volumetric rate
$Q_{\rm heat} = | \nabla \cdot {\bf F}_{\rm heat}|$)
must be included explicitly.
The key difference between the WTD and RLO paradigms, as described
in the previous section, rests on whether the heating is deposited
by fluctuations within an open flux tube or via impulsive injection
from other surrounding flux tubes.
Models can differ, of course, in the level of detail given in the
self-consistent heating rates.
Some descriptions of turbulence and magnetic reconnection specify
only the energy that is input into the system on the largest
spatial scales.
These models thus employ various phenomenological assumptions about
how that energy is eventually dissipated.
Other models follow the detailed microphysics of the dissipation
itself---usually by simulating either particle-particle collisions
or wave-particle interactions.
\item
{\em Additional momentum sources in the fast wind.}
In the 1970s and 1980s, it became increasingly evident that the
maximum mean plasma temperature in the open-field corona (i.e.,
approximately $(T_{p} + T_{e})/2)$ does not exceed
$\sim 2 \times 10^{6}$ K.
Even the measurements of higher values of $T_p$ in coronal
holes made in the 1990s did not substantially change this
constraint on the one-fluid average \citep[see, e.g.,][]{Ko06,Cr09}.
A solar wind with this mean temperature that is driven only by
gas pressure cannot accelerate to the highest speeds
(700--800 km s$^{-1}$) measured at 1 AU \citep{LH80}.
MHD fluctuations have been shown to exert an additional
``wave pressure'' or outward ponderomotive force on the mean fluid
\citep{D70,B71,OD98}.
This appears to be a necessary component of fast-wind models, and it
may even provide as much as half the acceleration in flux tubes
connected to polar coronal holes \citep[e.g.,][]{Cr04}.
\item
{\em A self-regulating mass flux.}
The Sun's mass loss rate, $\dot{M} \approx 2 \times 10^{-14}$
$M_{\odot}$ yr$^{-1}$, is generally believed to be
determined in the transition region by a balance between downward
heat conduction, local radiative losses, and the upward enthalpy
flux \citep{Ha82,Wi88,HL95,Le98}.
Self-consistent models should not artificially fix the properties
of the transition region and low corona.
Instead, models should allow these properties to ''float'' until a
natural, stable, and time-steady solution for the energy balance
is found.
\item
{\em Extended conduction and heating to 1 AU.}
Even far above the corona, {\em in situ} measurements show that
energy deposition and conductive energy transfer are still 
occurring \citep[e.g.,][]{Co68}.
The radial gradients of the proton and electron temperatures are
substantially shallower than would be the case if plasma parcels
were expanding adiabatically \citep{Ma83,Ma89,Ri95}.
{\em Helios} measurements of radial growth of the proton magnetic
moment between 0.3 and 1 AU \citep{SM83} point to specific
collisionless processes that continue to affect the energy budget
far from the Sun.
Recent empirical studies of the fast solar wind \citep{Br09,Cq09}
have shown that it is important to take careful account of electron
heat conduction in order to determine how the heating is partitioned
between protons and electrons.
Self-consistent models should be able to describe both the extended
heating that is observed and the transition from collisional to
collisionless conduction in the heliosphere.
\item
{\em Funnel-type magnetic field expansion.}
Most of the plasma that eventually is accelerated outwards as the
solar wind seems to originate in the lanes and vertices between
supergranular network cells in the chromosphere.
As height increases, the strong vertical magnetic field decreases
and the flux tubes expand laterally.
Thus, the flux tubes become magnetic funnels that eventually
merge with one another into a topologically complex ``canopy''
in the low corona \citep{Gb76,Do86,CvB05}.
Several semi-empirical studies of coronal heating and solar wind
acceleration show that this kind of funnel-like flux tube expansion
is necessary to producing realistic emission-line spectra in the
transition region \citep[e.g.,][]{Es05,Ma08,By08,Pu10}.
\end{enumerate}
The ways in which the above ingredients interact with one another
to produce a time-steady solar wind are complex and nonlinear.
Even though the wind flows upwards, sometimes the physics of the
extended corona (e.g., heating at heights of 1--2 solar radii above
the surface) can have a significant feedback on {\em lower} regions
in the atmosphere (e.g., frozen-in charge states that are set just
above the transition region).
It is important for a model to allow these various pieces of physics
to evolve together toward a stable steady state and not be
constrained by input assumptions.
For example, the lower boundary of the model should not be so high as
to exclude the transition region and upper chromosphere, and the upper
boundary should not be so low as to exclude the sonic or Alfv\'{e}nic
critical points of the flow.

\section{Wave/Turbulence Models}
\label{sec:3}

There has been substantial work over the last few decades devoted
to exploring the idea that coronal heating and solar wind acceleration
may be explained as a result of the dissipation of waves and
turbulent fluctuations.
No matter the relative importance of reconnections and loop-openings
in the low corona, we do know that waves and turbulent motions are
present everywhere from the photosphere to the heliosphere
\citep[see observational summaries of][]{TM95,CvB05,As08}.
Thus, it is of interest to determine how waves affect the mean
state of the plasma in the absence of any other sources of energy.

\citet{CvB07} described a set of WTD-type models in which the
time-steady plasma properties along a one-dimensional solar wind
flux tube are computed.
These model flux tubes are rooted in the optically thick solar
photosphere and are extended into interplanetary space.
The numerical code developed in that work, called ZEPHYR,
solves the one-fluid equations of mass, momentum,
and energy conservation simultaneously with transport
equations for Alfv\'{e}n and acoustic wave energy fluxes.
ZEPHYR is the first code capable of producing self-consistent
solutions for the photosphere, chromosphere, corona, and heliosphere
that combine:
(1) shock heating driven by an empirically guided spectrum of
acoustic waves, (2) extended heating from Alfv\'{e}n waves that
get partially reflected and then are dissipated by MHD turbulence,
and (3) wind acceleration from gradients of gas pressure,
acoustic wave pressure, and Alfv\'{e}n wave pressure.

The only input ``free parameters'' to the \citet{CvB07} models were
the photospheric lower boundary conditions for the waves and the
radial dependence of the magnetic field along the flux tube.
Photospheric measurements of the horizontal motions of intergranular
flux concentrations (i.e., G-band bright points having
$B \sim 1.5$ kG) were used to constrain the frequency spectrum
of Alfv\'{e}n waves at the lower boundary \citep[see also][]{Ni03}.
All models shown below used the same lower boundary conditions
and differed only in the rates of flux-tube superradial expansion.

The self-consistent coronal heating in the ZEPHYR models is the
result of propagating Alfv\'{e}n waves being partially reflected
by radial gradients in the density and magnetic field strength.
It has been shown that once there are {\em counter-propagating} wave
packets that interact with one another along a flux tube, a nonlinear
turbulent cascade can then occur relatively quickly \citep{Ir64,Kr65}.
The energy flux in the cascade, from large to small eddies,
terminates in dissipation and heating that extends from the low
corona out into the heliosphere
\citep[see also][]{Mt99,Dm02,CvB05,CH09,VV09,VV10,Cr10}.

In addition to the properties of the waves and turbulence, there
are other aspects of the models that determine how much coronal
heating occurs.
One of these is the radial location of the Parker ``critical point.''
This is the point at which the wind speed exceeds a critical speed
defined by the sound speed and the MHD wave amplitudes
\citep[see, e.g.,][]{J77}.
In models where the magnetic flux tubes expand purely radially,
there is usually just one unique location for this critical
point.
However, in flux tubes that undergo superradial expansion there
are multiple possible locations where the critical point could be
located.
These locations correspond to local minima in a potential-energy-like
quantity that was defined by \citet{KH76} and \citet{Vz03}.
However, only one of these points corresponds to a global energy
minimum, and thus only this one critical point location gives a
stable and time-steady solar wind.
For some models, changing the flux-tube expansion only slightly is
enough to alter the relative depths of these potential energy wells,
such that the critical point {\em shifts abruptly} from a
location close to the Sun to one much farther from the Sun.
When this occurs the global momentum and energy balance of the
solar wind changes abruptly as well.
Early studies \citep[e.g.,][]{LH80,Pn80} showed that high critical
points---where most of the heating is in the subsonic low
corona---correspond to dense and slow solar winds.
Conversely, low critical points---where the heating is mainly
in the supersonic outer corona---correspond to low-density and
fast solar winds.
For a ``stretched dipole'' model of the solar-minimum magnetic field,
the \citet{CvB07} models show precisely this kind of discontinuity 
between fast and slow winds at a heliospheric latitude of
$\sim 20^{\circ}$, similar to what {\em Ulysses} observed
\citep{Go96}.

\begin{figure}
\includegraphics[width=\textwidth]{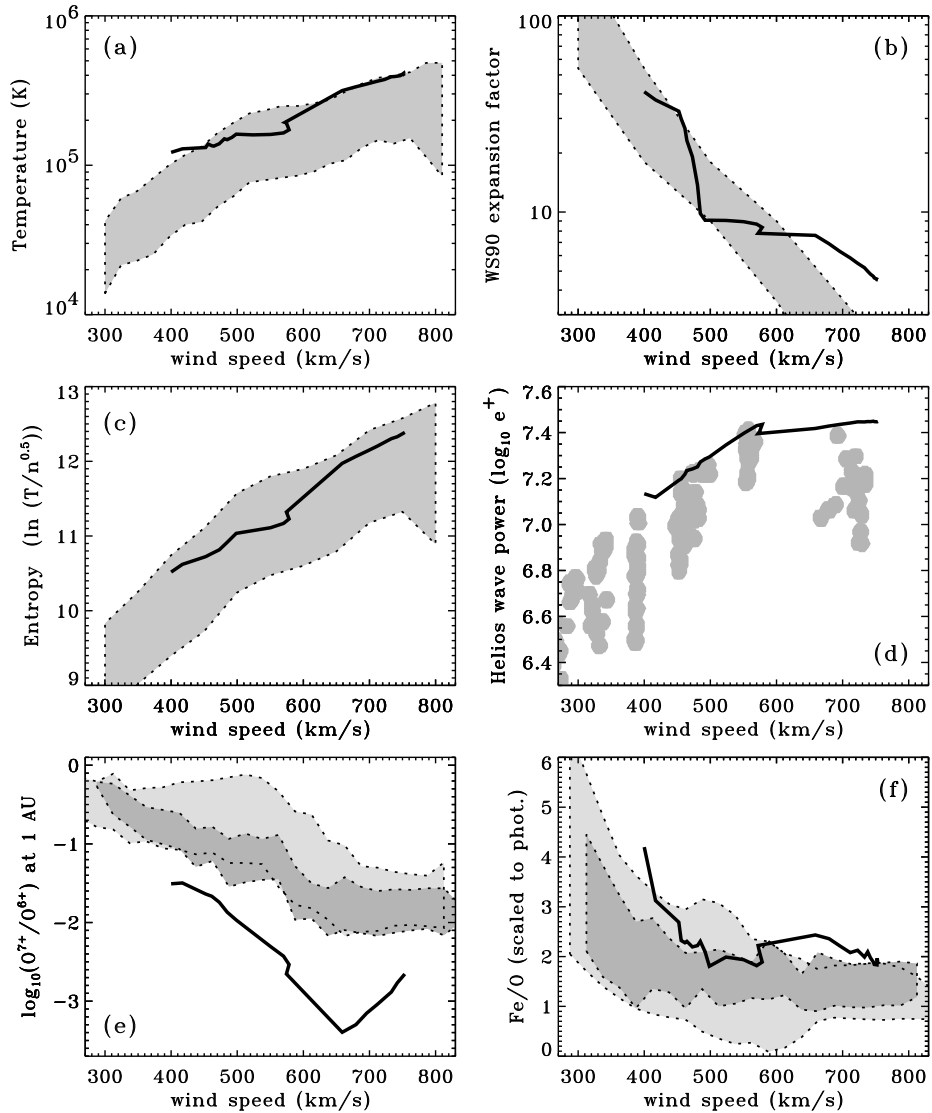}
\caption{Comparisons between the ZEPHYR models (thick black
curves) and observational data in the solar wind (gray regions).
See the main text for details \citep[see also][]{CvB07,Cr09}.
In all panels except (d), the gray regions show the data binned
by solar wind speed, showing only the regions within $\pm 1$
standard deviation of the mean value at each speed.}
\label{cranmer_fig1}
\end{figure}

Figure \ref{cranmer_fig1} shows several different comparisons between
solar wind measurements and the results of the \citet{CvB07} models.
In all six panels the observations and models are sorted by
the solar wind speed at 1 AU.
In panel (a), {\em in situ} proton temperature data are taken from
the {\em{ACE}}/SWEPAM online archive for the time periods between
1998 and 2005 \citep[see also][]{Mt06}.
Note that the plot juxtaposes the measured $T_p$ values with the
ZEPHYR one-fluid mean temperatures ($(T_{p} + T_{e})/2$), so
the comparison is not exact.
In panel (b), the coronal superradial expansion factor defined by
\citet{WS90} is shown as a function of wind speed, along with the
empirical trend of anticorrelation also demonstrated by \citet{WS90}.
In panel (c), {\em ACE} proton temperatures and densities have been
combined to form the specific entropy quantity that was found by
\citet{Pg04} to correlate strongly with wind speed.
In panel (d), the power in magnetic fluctuations measured by
{\em Helios} between 0.3 and 0.5 AU by \citet{Tu92} is compared with
a similar quantity estimated from the models in the way described by
\citet{CvB07}.

Panels (e) and (f) of Figure \ref{cranmer_fig1} show ion data from
{\em{Ulysses}}/SWICS taken at two different phases of that mission
(light gray: 1990--1994, dark gray: 1994--1995).
In panel (e), the measured ratio of O$^{7+}$ to O$^{6+}$ number
density is compared to ZEPHYR models of this ``frozen in''
nonequilibrium ionization state.
Although the overall trend with wind speed is similar to that in
the data, there is an overall shift downward in the modeled ratios.
This may be due to the fact that the models assume the electron
velocity distribution to be Maxwellian; i.e., the models ignore the
additional ionization that would be caused by a suprathermal
``electron halo'' in the corona \citep{EE00}.
In panel (f), the measured ratio of iron to oxygen elemental
abundances---normalized to their photospheric values---is compared
to models that apply the \citet{La04} idea for
first-ionization-potential (FIP) fractionation.
This idea utilizes the ponderomotive force exerted by Alfv\'{e}n
waves to accelerate ions in the partially ionized upper chromosphere,
while leaving the neutrals unaffected.
Note that the model results shown in panels (e) and (f) contradict
the commonly held assertion that slow-wind FIP and charge-state
properties can only be explained by the injection of plasma from
closed-field regions on the Sun \citep[see also][]{Pu10}.

Three other successful predictions of the ZEPHYR models are summarized
below:
\begin{enumerate}
\item
Recent {\em{Hinode}}/EIS measurements of strong Doppler shifts
at the edges of active regions indicate outflow speeds of order
100 km s$^{-1}$ in the coronal source regions of some slow wind
streams \citep{Ha08,Su10}.
As shown in Figure 7a of \citet{Cr10}, the ZEPHYR model of an
active-region-associated slow wind flux tube naturally exhibits
a local maximum in the outflow speed of the observed order of
magnitude at heights between 0.02 and 0.1 $R_{\odot}$ above the
photosphere.
\item
The original \citet{CvB07} model of the fast wind associated with
polar coronal holes used a magnetic field model consistent with
the 1996--1997 solar minimum.
However, the more recent 2008--2009 minimum has been seen to be
quite different.
\citet{Cr23} produced a new ZEPHYR model of the fast polar wind 
that used a lower magnetic field consistent with both solar-disk
and {\em in situ} measurements taken during 2008--2009.
The model produced changes in the plasma properties at 1 AU that
agree well with {\em Ulysses} measurements \citep[e.g.,][]{Mc08}.
For example, in both the models and the measurements, the
wind speed $u$ remains relatively unchanged, but the density $n$
and temperature $T$ decrease by factors of order 20\% and 10\%,
respectively.
The decreases in gas pressure (proportional to $nT$) and 
dynamic pressure (proportional to $n u^{2}$) are between 20\%
and 30\% for both the observations and models.
\item
The {\em{Helios}} probes measured Faraday rotation fluctuations
(FRFs) of polarized radio signals that passed through the solar
corona at heliocentric impact parameters between 2 and 15 $R_{\odot}$
in the ecliptic plane.
The magnitude of these fluctuations depends not only on the
amplitude of the Alfv\'{e}n waves in the corona, but also on the
density and the turbulent correlation length.
\citet{Ho10} compared the measured FRFs with predicted values from
ZEPHYR, and found excellent agreement when using the equatorial
streamer model originating at a colatitude of $28^{\circ}$.
\end{enumerate}

\section{Reconnection/Loop-Opening Models}
\label{sec:4}

It is clear from observations of the Sun's highly dynamical
``magnetic carpet'' that much of the coronal heating in closed-field
regions is driven by the interplay between the emergence, separation,
merging, and cancellation of small-scale flux tubes.
Magnetic reconnection seems to be the most likely channel for the
built-up magnetic energy to be converted to heat \citep[e.g.,][]{PF00}.
Thus, the RLO idea has a natural appeal since all open flux tubes
are rooted in the vicinity of closed loops \citep{Do86}.
In fact, isolated RLO-like reconnection events are already observed
in coronal holes as {\em polar jets} by {\em SOHO} and {\em Hinode}
\citep[e.g.,][]{Wa98,Sj07}.
Also, there are observed correlations between the lengths of
coronal loops, the electron temperature in the low corona, and the wind
speed at 1 AU \citep{Ge03} that are highly suggestive of a net transfer
of magnetic energy from the loops to the open-field regions
\citep[see also][]{Fi99,Fi03}.

Testing the RLO idea using theoretical models seems to be more difficult
than testing the WTD idea because of the complex multi-scale nature
of the relevant magnetic topology.
It could be argued that one needs to create three-dimensional and
time-dependent models of the magnetic carpet in order to fully
take account of all interactions between the closed and open flux systems.
Several key questions remain to be answered.
For example, how much of the magnetic energy that is liberated by
reconnection goes into simply reconfiguring the closed fields, and
how much goes into changing closed fields into open fields?
Specifically, what is the actual {\em rate} at which magnetic flux
opens up in the magnetic carpet?
Can the observed polar jets provide enough energy to drive a
significant fraction of the solar wind?
Lastly, how is the reconnection energy distributed into various
forms (e.g., bulk kinetic energy, thermal energy, waves, or
energetic particles) that can each affect the accelerating wind in
different ways?

Recent work has started to provide tentative answers to some of the
above questions.
\citet{CvB10} developed Monte Carlo simulations of the
time-varying magnetic carpet and its connection to the large-scale
coronal field.
These models were constructed for a range of different magnetic
flux imbalance ratios---i.e., for both quiet regions and coronal
holes.
The models agree with observed emergence rates, surface flux
densities, and number distributions of magnetic elements.
Despite having no imposed supergranular motions in the models,
a realistic network of magnetic ``funnels'' appeared spontaneously
as the result of diffusion-limited aggregation from smaller
magnetic concentrations.
\citet{CvB10} computed the rates at which closed field lines open
up (i.e., the recycling times for open flux), and they estimated
the energy fluxes released in reconnection events that involve the
opening up of closed flux tubes.
For quiet regions and mixed-polarity coronal holes, these energy
fluxes were found to be significantly smaller than those required to
accelerate the solar wind.
In other words, only a tiny fraction of the Poynting flux delivered
into the corona by emerging bipoles seems to be released via magnetic
reconnection in RLO-like events.
On the other hand, for the most imbalanced (i.e., unipolar) coronal
holes, the energy in flux-opening events may be large enough to
power the solar wind.
However, in those cases the overall recycling times are far longer
than the time it takes the solar wind to accelerate up into the
low corona.
Thus, RLO processes on supergranular scales may be responsible for
the intermittent jets in coronal holes, but probably not for the
majority of the ``bulk'' solar wind acceleration.

\section{Conclusions}
\label{sec:5}

Despite recent progress made with both the WTD and RLO approaches to
plasma heating and acceleration, we still do not have conclusive
answers about whether one idea or the other is dominant in the
actual corona and solar wind.
It is possible, of course, that qualitatively different mechanisms
may govern the heating and acceleration in different types of
solar wind streams.
It is also possible that some kind of combination of the WTD
and RLO paradigms may be more valid than either idea in isolation.

An important next step in the process is to determine what specific
observations can best test these ideas, with the goal of convincingly
verifying and/or falsifying them.
Also, a related future step will be to build three-dimensional models
of the Sun-heliosphere system that include WTD and RLO physics as
their ``coronal heating functions.''
These kinds of simulations can be customized for specific time periods,
and be used to make straightforward comparisons with various kinds
of existing remote-sensing and {\em in situ} measurements.
Some recent progress in producing computationally efficient
approximations to the rates of WTD wave reflection and heating has
been reported by \citet{CH09} and \citet{Cr10}.

In addition to expanding the scope of the models, it will also be
important to develop a better understanding of the physics of MHD
wave generation, propagation, and dissipation.
Recent observations of Alfv\'{e}n waves in the complex lower
atmosphere indicate that the energy in fluctuations is distributed
intermittently between spicules, loops, and the open-field
corona \citep{DP07,Tz07,Tz09,FT09}.
Magnetic flux tubes that thread the upper chromosphere and low
corona support a wide range of possible nonlinear couplings between
compressible and incompressible modes \citep[e.g.,][]{Bg03,Hs05}.
For example, acoustic waves from non-magnetic regions of the
chromosphere may encounter a swaying flux tube and be converted
into Alfv\'{e}n or fast-mode waves as they pass through the tube.
The study of energy transport across small-scale {\em interfaces}
in the solar atmosphere may be crucial to producing more accurate
and predictive models.

Finally, it is important for future models to take account of the
kinetic and multi-fluid nature of coronal heating and solar
wind acceleration \citep{Ma06}.
In the WTD paradigm, for example, a description of the large-scale
energy flux injected into a turbulent cascade is only the
first chapter in the story.
Better descriptions of the anisotropic cascade process, the eventual
kinetic dissipation of the fluctuations, and the subsequent
energization of electrons, protons, and heavy ions are needed to
complete the picture.
Remote-sensing measurements of strong preferential heating and
acceleration for heavy ions (e.g., O$^{+5}$) in coronal holes have
spurred a great deal of theoretical work in this direction
\citep[see, e.g.,][]{HI02,Cr02,Cr09}.
A proper accounting of these kinetic effects will lead to
more concrete predictions for measurements to be made by space
missions such as {\em Solar Probe} \citep{Mc07}
and {\em Solar Orbiter} \citep{MF07},
as well as next-generation ultraviolet coronagraph spectroscopy
that could follow up on the successes of the UVCS instrument
on {\em SOHO} \citep{Ko06,Ko08}.

\begin{acknowledgements}
The author thanks the organizers and participants of the
{\em ``Multi-Scale Physics in Coronal Heating and Solar Wind
Acceleration''} Workshop of the International Space Science
Institute (ISSI) for making possible such an extremely
fruitful gathering of theorists, observers, and experimenters.
This work was supported by the National Aeronautics and Space
Administration (NASA) under grants {NNG\-04\-GE77G,}
{NNX\-09\-AB27G,} and {NNX\-10\-AC11G}
to the Smithsonian Astrophysical Observatory.

\end{acknowledgements}

\end{document}